# The domestic localization of knowledge flows as evidenced by publication citation: The case of Italy[*1]

*Giovanni Abramo (corresponding author)*

Laboratory for Studies in Research Evaluation, Institute for System Analysis and Computer Science (IASI-CNR), National Research Council of Italy
giovanni.abramo@uniroma2.it

*Ciriaco Andrea D'Angelo*

Department of Engineering and Management, University of Rome "Tor Vergata" and Laboratory for Studies in Research Evaluation, Institute for System Analysis and Computer Science (IASI-CNR), Italy
dangelo@dii.uniroma2.it

**Abstract**

This work applies a new approach to measure knowledge flows. Assuming that citation linkages between articles imply a flow of knowledge from the cited to the citing authors, we investigate the geographic flows of scientific knowledge produced in Italy across its regions, at both overall and field level. Furthermore, we measure the the specialization indexes for outflows and inflows of knowledge by a given region. Findings show that larger regions in terms of research output are more likely net exporters of new knowledge. At the same time, we register a positive correlation between the share of intraregional flows and the size of overall scientific output of a region.

---

# 1. Introduction

The 2018 Nobel Prize in economics was awarded to Paul Romer for his contributions to the theory of long-run economic growth with "endogenous" technological change (Romer, 1986; 1990). Romer's endogenous growth theory ties the development of new ideas to the number of people working in the knowledge sector (i.e. an effort devoted to R&D). These new ideas make everyone else producing regular goods and services more productive – that is, ideas increase total factors productivity. It concurs to that a peculiar feature of ideas, the fact that they are "non-rival" (meaning that one's use of an idea, like a recipe or a mathematical formula, does not prevent somebody's else use of it). In theory, public knowledge (as that encoded in publications) can be shared endlessly. However, in practice the diffusion of knowledge decays with the geographical distance from the idea generator. Since knowledge transfer cannot be observed directly (Jaffe, Trajtenberg, & Fogarty, 2000), one relies on proxy measures, notably citations. Jaffe, Trajtenberg, and Henderson (1993) compared the geographic location of patent citations with that of the cited patents, to investigate the extent to which knowledge flows are geographically localized. They found that citations to domestic patents are more likely to be domestic, and more likely to come from the same state as the cited patents.

Assuming that citation linkages between articles imply a flow of knowledge from the cited to the citing authors (Mehta, Rysman, & Simcoe, 2010; Van Leeuwen & Tijssen, 2000), few scholars relied on publication citations to investigate the international geographic flows of scientific knowledge. Rabkin, Eisemon, Lafitte-Houssat, and McLean Rathgeber (1979) explored world visibility for four departments (botany, zoology, mathematics, and physics) of the universities of Nairobi (Kenya) and Ibadan (Nigeria). At the level of the single field, Stegmann and Grohmann (2001) measured knowledge "export" in the Dermatology & Venereal Diseases category of the 1996 CD-ROM Journal Citation Reports (JCR), and in seven dermatology journals not listed in the 1996 JCR. Hassan and Haddawy (2013) mapped knowledge flows from the United States to other countries in the field of Energy over the years 1996-2009. Abramo and D'Angelo (2018) trailed international flows of knowledge produced in Italy, in over 200 fields, by analysing publication citations. Abramo, D'Angelo, and Carloni (2018) conceptualized the "balance of knowledge flows" (BKF) at the international level. Among others, the authors measured the share of domestic vs foreign flows generated by a country's research system, by field and as compared to other countries.

Abramo and D'Angelo (2019) were the first ever to trail the domestic flows of knowledge by publication citations, in Italy. This study expands the above one to include the calculation of the regions' specialization indexes. After showing the spread of the flows of knowledge across regions, we measure the specialization indexes for outflows (export) and inflows (import) of knowledge by a given region. Specialization indexes measure the extent to which a region's knowledge flows differ from those of the rest of the country. In simple terms, they measure a region's capacity to "export" knowledge to other regions, or to "import" knowledge from other regions, as compared to the rest of the country, across all research fields.

We use a bibliometric approach, assuming that all new knowledge produced is measured by publications indexed in bibliographic repertories. We also assume that citations are proxies of scholarly impact, i.e. when a publication is cited it has had an impact on scientific advancement because other scholars have drawn on it, more or less heavily, for the further advancement of science. All limitations and assumptions typical of bibliometric analyses then apply. As for publications, it must me noted that not all new knowledge is encoded in publications (e.g. tacit knowledge), and not all publications are indexed in bibliographic repertories. Furthermore, stating that citations certify knowledge flows does not imply that there are no exceptions, rather that it is the norm. Citations in fact are not always certification of real flows and representative of all flows. Uncitedness, undercitation, and overcitation may actually occur. Finally, citation-based analysis is unable to capture flows outside the scientific system, such as that of practitioners (e.g. a physician applying a new pharmacological protocol after reading relevant literature), students, or industry.[2]

This work intends to measure the flows of knowledge[3] across regions. Identifying the administrative regions of all world institutions publishing in a period of time is a formidable task. For that reason, we restrict our analysis to the national level, and in particular to the authors' country, Italy. While results cannot be generalized to other countries, the study can be easily replicated in other national contexts, and provide useful information to the policy maker, such as the share of intra- vs extra-regional knowledge flows generated by a region's research system, by field and as compared to other regions. A very high intra-regional share is expected in those fields where research is context specific or mainly oriented towards intra-regional needs.

We observe Italian publications indexed in Web of Science (WoS) in 2010-2012, and their citing domestic publications up to 31/05/2017,[4] to measure the outflows of knowledge produced in a region to other regions in Italy, and the inflows of knowledge produced by other regions in a region. The latter allows us to set up a region's balance of knowledge flows (RBKF), which would register a surplus when the difference between knowledge outflows and inflows is positive, a deficit when the opposite is true. We measure also the share of intra- vs extra-regional flows within each of the 20 administrative regions in Italy.[5]

In the next section we present the data and method of analysis. Section 3 provides the results of the analysis of knowledge flows across regions, both at overall and at field level. Section 4 presents the measurements of the specialization indexes at region and field level. Section 5 closes the work with our considerations on the relevance of the study.

## 2. Data and method

In the period 2010-2012, the seven main databases of the WoS *core collection*[6] indexes 255399 publications showing Italian affiliations.

To measure the regional outflows of knowledge one needs to identify the region of production of both the citing and cited publications. Because of increasing research collaboration at both national and international level, identifying the region of production of a publication may reveal not so straightforward. Various approaches for assigning an inter-regional authored publication to a region can be envisaged: i) to each region the institutions in the address list belong; ii) to one single region, based on the frequency the authors of that region (or the institutions of that region), occur in the address list; or based on the affiliation of the corresponding author, or first and last authors in non-alphabetically ordered bylines; iii) fractionalizing the publication by the number of regions, institutions or authors.

---

[2] We refer the reader to Abramo (2018) for a thorough discussion on the subject.
[3] In this work, we use the term knowledge flows without any further distinction, such as between horizontal and vertical knowledge flows.
[4] We include self-citing publications, because it may not matter whether the subsequent development that flows from a publication is performed by the same author(s), as long as it is performed in her or his region.
[5] The spillovers we measure do not account for the sharing of knowledge among co-authors inherent in any research collaboration.
[6] SCI-E: Science Citation Index Expanded; SSCI: Social Sciences Citation Index; A&HCI: Arts & Humanities Citation Index; CPCI-S: Conference Proceedings Citation Index- Science; CPCI-SSH: Conference Proceedings Citation Index- Social Science & Humanities; BKCI-S: Book Citation Index– Science; BKCI-SSH: Book Citation Index– Social Sciences & Humanities.

The convention we adopt here is the following: We define a publication as "made in" a region if at least 50% of its co-authors are affiliated to organizations located in that region. Because we are dealing with domestic knowledge flows, we had to exclude publications produced abroad, i.e. those with more than half co-authors affiliated to foreign institutions. Furthermore, 17401 publications are multi-authored and lack the author-affiliation link. We were forced to exclude those as well from the analysis.

The final dataset of analysis is then composed of 167630 "made in" Italy publications (Table 1). Of them, 163395 are "made in" single regions only, and 4235 equally "made in" two regions, meaning that 50% of the authors are affiliated to institutions located in the first region and 50% in the second. To identify and assign the region of production to publications, we have developed a matching application of geographic metadata ("affil_CITY", "affil_PROVINCE", and "affil_ZIP_CODE"), as reported in the address of publications. We have run the same application also to identify the region(s) of production of the citing publications.

To account for multiple affiliations of authors, we adopt a fractional counting method. In case of authors with $m$ different affiliations, we assign $1/m$ to each of her or his bibliometric addresses. Of course, the issue reveals critical for authors affiliated to institutions located in different countries and/or regions. To exemplify, consider the publication with WoS code 000309458000001.

> **Nonlinear dynamics of beta-induced Alfven eigenmode driven by energetic particles**
>
> By: Wang, X (Wang, X.)[1]; Briguglio, S (Briguglio, S.)[2]; Chen, L (Chen, L.)[1,3]; Di Troia, C (Di Troia, C.)[2]; Fogaccia, G (Fogaccia, G.)[2]; Vlad, G (Vlad, G.)[2]; Zonca, F (Zonca, F.)[1,2]
>
> **Author Information**
> Reprint Address: Wang, X (reprint author)
>    Zhejiang Univ, Inst Fus Theory & Simulat, Hangzhou 310027, Zhejiang, Peoples R China.
> Addresses:
>    [ 1 ] Zhejiang Univ, Inst Fus Theory & Simulat, Hangzhou 310027, Zhejiang, Peoples R China
>    [ 2 ] Assoc Euratom ENEA Fus, I-00044 Frascati, Italy
>    [ 3 ] Univ Calif Irvine, Dept Phys & Astron, Irvine, CA 92697 USA

Its seven co-authors are affiliated to three different institutions, located in China, Italy and the U.S. Chen L. and Zonca F. show double affiliation. The Italian institution ENEA-Euratom (localized in the region Latium), scores 4.5 authorships out of 7, because 4 authors (Briguglio S., Di Troia C., Fogaccia G., Vlad G.) are affiliated solely to it and one, Zonca F., is affiliated also to another institution. Therefore, the publication is defined as "made in" Latium. To measure inter-regional flows, we replicate to the regional level the approach detailed in Abramo & D'Angelo (2018). When a publication is cited it has given rise to a "benefit". The number of "benefits" deriving from a publication equals the number of citations, and if the citing publication is co-authored by scholars from one or more regions, the benefit has crossed a regional administrative boundary. In the case of a citing publication whose address list shows institutions located in *p* different regions, the same benefit (citation) is "gained" contemporaneously by *p* regions, so we can say that it has given rise to *p* equal "gains", one for each region of the Italian institutions listed in the affiliation list of the citing publication. A publication cited by *q* other publications would give rise to *q* benefits and *q* x *p* gains. Among the *p* citing regions there could be also the region the cited publication is made in. In this case we define the relevant gain as "intra-regional".

To exemplify, consider the publication with WoS code 000209048200010:

> Chinski, A., Foltran, F., Gregori, D., Passali, D., & Bellussi, L. (2011). Foreign bodies in the ears in children: The experience of the buenos aires pediatric ORL clinic. *Turkish Journal of Pediatrics, 53*(4), 425-429.

Such publication is classified as "made in" Tuscany since three of its five co-authors (Foltran, F.; Passali, D.; Bellussi, L.) belong to two institutions (University of Siena and University of Pisa) located in that region. At 31/05/2017 the publication has accrued nine citations, 6 of which by publications with at least one Italian address as detailed below.

> Foltran, F., Passali, F. M., Berchialla, P., Gregori, D., Pitkäranta, A., Slapak, I., . . . Raine, C. (2012). Toys in the upper aerodigestive tract: New evidence on their risk as emerging from the susy safe study. *International Journal of Pediatric Otorhinolaryngology, 76*(SUPPL. 1), S61-S66.
> *Italian regions associated to authors' affiliations*: Friuli Venezia Giulia; Latium; Piedmont; Tuscany; Veneto

> Gregori, D., et al. (2012). The susy safe project overview after the first four years of activity. *International Journal of Pediatric Otorhinolaryngology, 76*(SUPPL. 1), S3-S11.
> *Italian regions associated to authors' affiliations*: Campania; Emilia Romagna; Friuli Venezia Giulia; Latium; Piedmont; Tuscany; Veneto

> Moretti, C., & Foltran, F. (2012). Prevention and early recognition: The role of family pediatrician. *International Journal of Pediatric Otorhinolaryngology, 76*(SUPPL. 1), S39-S41.
> *Italian regions associated to authors' affiliations*: Veneto

> Sarafoleanu, C., Ballali, S., Gregori, D., Bellussi, L., & Passali, D. (2012). Retrospective study on romanian foreign bodies injuries in children. *International Journal of Pediatric Otorhinolaryngology, 76*(SUPPL. 1), S73-S75.
> *Italian regions associated to authors' affiliations*: Friuli Venezia Giulia; Tuscany; Veneto

> Sih, T., Bunnag, C., Ballali, S., Lauriello, M., & Bellussi, L. (2012). Nuts and seed: A natural yet dangerous foreign body. *International Journal of Pediatric Otorhinolaryngology, 76*(SUPPL. 1), S49-S52.
> *Italian regions associated to authors' affiliations*: Abruzzo; Friuli Venezia Giulia; Tuscany

> Slapak, I., Passali, F. M., Passali, G. C., Gulati, A., Gregori, D., Foltran, F., . . . Raine, C. (2012). Non food foreign body injuries. *International Journal of Pediatric Otorhinolaryngology, 76*(SUPPL. 1), S26-S32.
> *Italian regions associated to authors' affiliations*: Latium

In brief, this publication generates six domestic benefits and 20 domestic gains, four of which intra-regional (Tuscany-Tuscany).

The RBKF is constructed for each region measuring the gains associated to the inflows and outflows of knowledge among the 20 Italian regions.

The overall publications in the 2010-2012 period, and the relevant benefits and gains per Italian region are shown in Table 1. The first row shows, for example, that researchers from Abruzzo in the three-year under observation authored 6541 publications, 2856 of which "made in", since at least 50% of their coauthors listed in the byline are affiliated to institutions located in that region. In turn, 72.1% of said publications are cited (at 31/05/2017) generating a total of 8552 domestic benefits, with an average of 4.15 domestic benefits per "made in" cited publication (8552/2060). On average 1.63 regions appropriate such benefits for a total of 13973 gains (8552*1.63), of which 39.6% intraregional, i.e. related to citing publications authored by other researchers from the Abruzzo region. The share of intraregional gains varies from a minimum of 24% in the smallest region, Valle D'Aosta, to a max of 54.4% in Sicily, one of the two island regions.

In general, there is a positive correlation (Spearman $\rho = 0.608$) between the share of intraregional gains and the size of overall scientific production of a region. This can be due to the fact that in (scientifically) large regions it is likely to find large research laboratories/groups conducting research on topics of common interest.

*Table 1: 2010-2012 publications, citations, benefits and gains at 31/05/2017, by region*

| Region | Total publications | "Made in" publications | Of which cited (a) | Total domestic benefits (No. of citations from Italian regions) (b) | Average domestic benefits per cited publication (b/a) | Total domestic gains (c) | Of which intra-regional | Average domestic gains per benefit (c/b) |
|---|---|---|---|---|---|---|---|---|
| Abruzzo | 6541 | 2856 (43.7%) | 2060 (72.1%) | 8552 | 4.15 | 13973 | 5529 (39.6%) | 1.63 |
| Basilicata | 1858 | 758 (40.8%) | 582 (76.8%) | 2384 | 4.10 | 3637 | 1567 (43.1%) | 1.53 |
| Calabria | 5581 | 2962 (53.1%) | 2258 (76.2%) | 10443 | 4.62 | 15165 | 7751 (51.1%) | 1.45 |
| Campania | 21365 | 12632 (59.1%) | 9134 (72.3%) | 45305 | 4.96 | 64386 | 34349 (53.3%) | 1.42 |
| Emilia Romagna | 31773 | 16491 (51.9%) | 11780 (71.4%) | 51856 | 4.40 | 77055 | 36692 (47.6%) | 1.49 |
| Friuli Venezia Giulia | 11907 | 4830 (40.6%) | 3549 (73.5%) | 15812 | 4.46 | 23549 | 10500 (44.6%) | 1.49 |
| Latium | 48519 | 26485 (54.6%) | 18643 (70.4%) | 79858 | 4.28 | 120646 | 58770 (48.7%) | 1.51 |
| Liguria | 10808 | 5054 (46.8%) | 3516 (69.6%) | 14660 | 4.17 | 22325 | 10197 (45.7%) | 1.52 |
| Lombardy | 56236 | 31311 (55.7%) | 21543 (68.8%) | 95641 | 4.44 | 140911 | 70500 (50.0%) | 1.47 |
| Marche | 5740 | 2960 (51.6%) | 2150 (72.6%) | 9040 | 4.20 | 14037 | 6042 (43.0%) | 1.55 |
| Molise | 1529 | 377 (24.7%) | 288 (76.4%) | 1286 | 4.47 | 2347 | 771 (32.9%) | 1.83 |
| Piedmont | 20437 | 11015 (53.9%) | 7731 (70.2%) | 33636 | 4.35 | 47614 | 23559 (49.5%) | 1.42 |
| Puglia | 13106 | 6923 (52.8%) | 5060 (73.1%) | 21859 | 4.32 | 32242 | 15386 (47.7%) | 1.47 |
| Sardinia | 6009 | 2959 (49.2%) | 2198 (74.3%) | 9202 | 4.19 | 13369 | 6605 (49.4%) | 1.45 |
| Sicily | 16341 | 9722 (59.5%) | 7306 (75.1%) | 34648 | 4.74 | 48414 | 26327 (54.4%) | 1.40 |
| Tuscany | 31204 | 16894 (54.1%) | 11767 (69.7%) | 53209 | 4.52 | 78646 | 38210 (48.6%) | 1.48 |
| Trentino Alto Adige | 6224 | 2969 (47.7%) | 2178 (73.4%) | 9313 | 4.28 | 12875 | 6704 (52.1%) | 1.38 |
| Umbria | 5560 | 2477 (44.6%) | 1913 (77.2%) | 10233 | 5.35 | 15469 | 7217 (46.7%) | 1.51 |
| Valle D'Aosta | 184 | 53 (28.8%) | 36 (67.9%) | 91 | 2.53 | 179 | 43 (24.0%) | 1.97 |
| Veneto | 24158 | 12137 (50.2%) | 8675 (71.5%) | 37291 | 4.30 | 54164 | 26139 (48.3%) | 1.45 |

# 3. Results and analysis

## 3.1 The regional balance of knowledge flows at overall level

The RBKF of all 20 Italian regions is shown Table 2. For the sake of easier reading, we present the results concerning the Latium region. As shown in Table 1, the "made in" Latium cited publications are 18643. Such publications generate a total of 120646 gains, 58770 of which appropriated by Latium institutions. The remaining 61876 (as shown in column 2 of Table 2) are appropriated by the other 19 regions, which in turn publish altogether 84693 cited publications, generating a total of 352283 domestic gains, of which Latium appropriates 55862 (15.9%). The Latium RBKF is therefore positive (surplus) and equal to +6014, given the imbalance between the generated domestic gains (61876) and the earned gains (55862). The RBKF is also positive for Lombardy (+17386), Tuscany (+4738), Piedmont (+1710) and Campania (+802), i.e. the largest regions in terms of scientific production. The only exception among the largest regions is Emilia Romagna showing a negative RBKF (-1621), as the remaining 14 regions. In general, there occurs a positive correlation (Spearman $\rho = 0.849$) between the size of scientific production of a region and the value of its RBKF.

*Table 2: The regional balance of knowledge flows (RBKF). Among parenthesis percentages out of total gains*

| Region | Extra-regional gains generated (a) | Cited extra-regional publications | Extra-regional gains generated by extra-regional publications | Earned gains (b) | RBKF (a-b) |
|---|---|---|---|---|---|
| Abruzzo | 8444 (60.4%) | 97992 | 394856 | 13289 | -4845 |
| Basilicata | 2070 (56.9%) | 99210 | 404337 | 3808 | -1738 |
| Calabria | 7414 (48.9%) | 97799 | 398397 | 9748 | -2334 |
| Campania | 30037 (46.7%) | 92021 | 378910 | 29235 | +802 |
| Emilia Romagna | 40363 (52.4%) | 90121 | 366161 | 41984 | -1621 |
| Friuli Venezia Giulia | 13049 (55.4%) | 96802 | 391831 | 16314 | -3265 |
| Latium | 61876 (51.3%) | 84693 | 352283 | 55862 | +6014 |
| Liguria | 12128 (54.3%) | 96891 | 393656 | 14489 | -2361 |
| Lombardy | 70411 (50.0%) | 82326 | 355120 | 53025 | +17386 |
| Marche | 7995 (57.0%) | 97951 | 398897 | 9248 | -1253 |
| Molise | 1576 (67.1%) | 99457 | 403568 | 4577 | -3001 |
| Piedmont | 24055 (50.5%) | 93468 | 385800 | 22345 | +1710 |
| Puglia | 16856 (52.3%) | 95506 | 389875 | 18270 | -1414 |
| Sardinia | 6764 (50.6%) | 97892 | 399111 | 9034 | -2270 |
| Sicily | 22087 (45.6%) | 93610 | 385597 | 22548 | -461 |
| Tuscany | 40436 (51.4%) | 90236 | 372447 | 35698 | +4738 |
| Trentino Alto Adige | 6171 (47.9%) | 97999 | 399015 | 9130 | -2959 |
| Umbria | 8252 (53.3%) | 98079 | 397977 | 10168 | -1916 |
| Valle D'Aosta | 136 (76.0%) | 99683 | 407630 | 515 | -379 |
| Veneto | 28025 (51.7%) | 92716 | 379287 | 28858 | -833 |

Figure 1 shows the data of columns 2, 5 and 6 of Table 2 for a better overview of the entity of generated and earned gains by each region. Regions are ordered by total flows (both inflows and outflows). Four out of the top five regions show a positive RBKF (Lombardy, Latium, Tuscany and Campania), and one (Emilia Romagna) negative. With the sole exception of Piedmont, all the others have a negative RBKF.

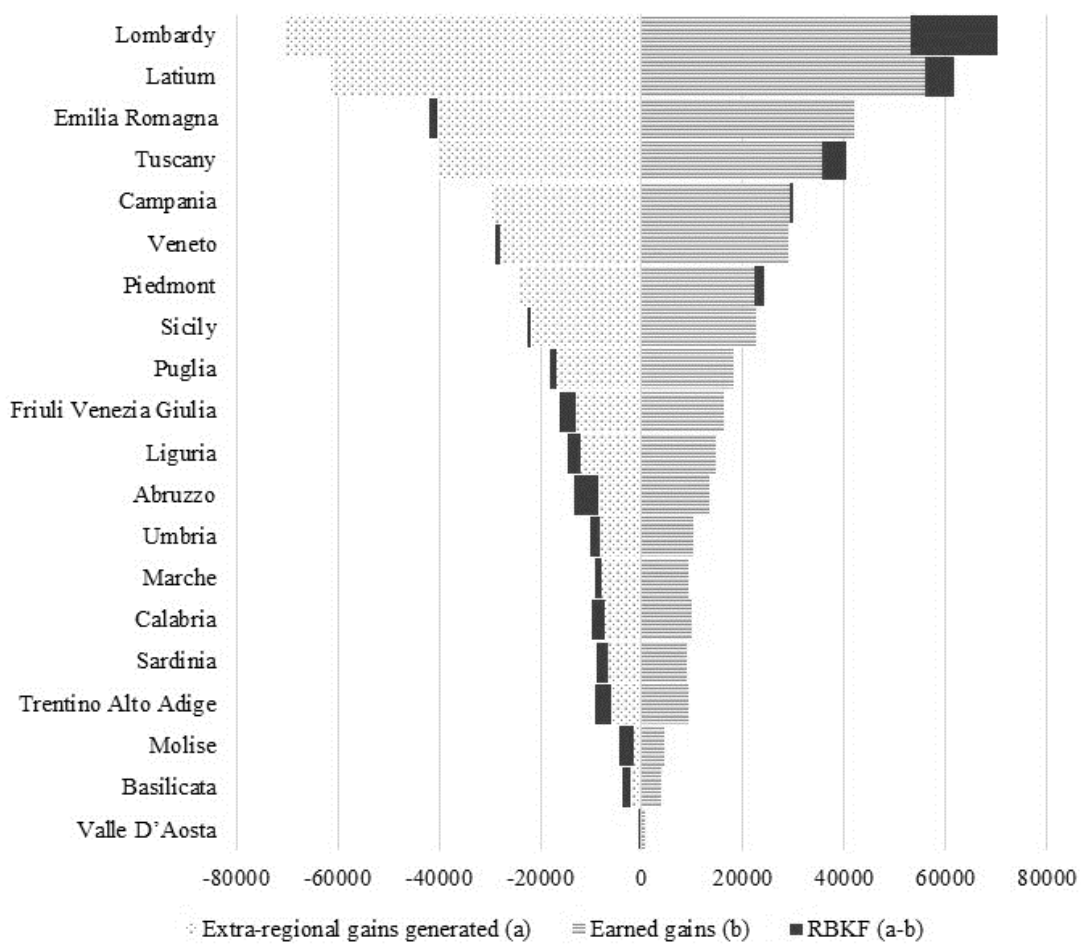

*Figure 1: Regional balance of knowledge flows (RBKF), for each Italian region*

Table 3 shows inter-regional knowledge flows. Data on the main diagonal of the matrix illustrate the share of gains generated by a region which remain within that region (intraregional gains).[7]

The matrix should be read by row, because the row vector shows the regional flows of knowledge produced in a given region (summing up to 100%). To exemplify, 16.2% of knowledge flows out of Abruzzo are appropriated by Latium, double as much as by Lombardy (8.0%), followed by Emilia Romagna (5.7%) and Tuscany (4.5%).

Needless to say, the matrix may be read by column also, in which case it will show an insight into the origin of the knowledge inflows of a given region (of course, in this case values should be rescaled, since the column does not sum up to 100%).

---

[7] The same as in column 8 of Table 1.

*Table 3: Overall import-export of knowledge (percentages out of total gains) between regions*

| | | Earning | | | | | | | | | | | | | | | | | | | |
|---|---|---|---|---|---|---|---|---|---|---|---|---|---|---|---|---|---|---|---|---|---|
| | Region | Abruzzo | Basilicata | Calabria | Campania | Emilia Romagna | Friuli Venezia Giulia | Latium | Liguria | Lombardy | Marche | Molise | Piedmont | Puglia | Sardinia | Sicily | Tuscany | Trentino Alto Adige | Umbria | Valle D' Aosta | Veneto |
| Generating | Abruzzo | 39.6 | 0.3 | 0.7 | 4.1 | 5.7 | 1.7 | 16.2 | 1.3 | 8.0 | 1.6 | 0.7 | 2.2 | 3.1 | 0.9 | 3.0 | 4.5 | 0.4 | 2.8 | 0.0 | 3.4 |
| | Basilicata | 1.3 | 43.1 | 2.0 | 9.7 | 4.1 | 2.0 | 8.0 | 0.7 | 4.4 | 0.9 | 0.8 | 1.4 | 7.7 | 1.0 | 4.0 | 3.8 | 0.7 | 2.5 | 0.1 | 1.8 |
| | Calabria | 1.3 | 0.7 | 51.1 | 5.9 | 4.4 | 1.0 | 7.9 | 2.5 | 5.1 | 0.9 | 0.3 | 1.7 | 2.0 | 1.4 | 6.6 | 3.7 | 0.6 | 1.1 | 0.0 | 1.8 |
| | Campania | 1.6 | 1.1 | 1.6 | 53.3 | 4.1 | 1.3 | 8.5 | 1.5 | 5.9 | 1.1 | 1.0 | 2.1 | 2.6 | 0.9 | 3.3 | 5.6 | 0.6 | 1.4 | 0.0 | 2.3 |
| | Emilia Romagna | 1.5 | 0.3 | 0.8 | 3.3 | 47.6 | 2.6 | 8.5 | 1.8 | 9.6 | 1.4 | 0.3 | 3.1 | 2.2 | 1.1 | 3.1 | 5.1 | 1.1 | 1.1 | 0.0 | 5.4 |
| | Friuli Venezia Giulia | 1.5 | 0.3 | 0.9 | 3.0 | 6.5 | 44.6 | 8.0 | 1.5 | 8.6 | 1.5 | 0.2 | 3.0 | 2.3 | 0.9 | 2.7 | 5.2 | 1.9 | 0.9 | 0.1 | 6.7 |
| | Latium | 3.7 | 0.6 | 1.4 | 4.7 | 5.6 | 2.2 | 48.7 | 1.7 | 8.4 | 1.3 | 1.1 | 2.6 | 2.5 | 1.3 | 3.1 | 5.2 | 0.8 | 1.6 | 0.0 | 3.4 |
| | Liguria | 1.2 | 0.1 | 1.7 | 4.0 | 5.5 | 1.6 | 7.2 | 45.7 | 9.8 | 1.5 | 0.3 | 3.4 | 2.5 | 1.3 | 2.6 | 5.6 | 1.2 | 0.8 | 0.1 | 4.0 |
| | Lombardy | 1.2 | 0.3 | 1.0 | 3.6 | 6.2 | 2.3 | 8.3 | 2.5 | 50.0 | 0.9 | 0.5 | 4.2 | 2.5 | 1.0 | 2.8 | 5.1 | 1.4 | 1.3 | 0.1 | 4.6 |
| | Marche | 1.7 | 0.8 | 1.1 | 4.3 | 7.1 | 2.1 | 9.4 | 1.6 | 7.1 | 43.0 | 0.5 | 2.3 | 2.1 | 1.5 | 2.5 | 5.5 | 1.2 | 2.5 | 0.0 | 3.7 |
| | Molise | 1.4 | 1.0 | 1.6 | 11.0 | 4.3 | 0.9 | 16.7 | 1.0 | 8.1 | 0.8 | 32.9 | 1.5 | 3.8 | 1.9 | 2.9 | 4.0 | 1.7 | 1.3 | 0.1 | 3.0 |
| | Piedmont | 1.0 | 0.3 | 0.9 | 3.0 | 5.9 | 1.8 | 6.9 | 1.9 | 10.3 | 1.1 | 0.3 | 49.5 | 2.1 | 1.2 | 2.9 | 4.2 | 1.5 | 1.0 | 0.3 | 4.0 |
| | Puglia | 1.5 | 1.8 | 1.7 | 4.4 | 5.3 | 1.7 | 7.9 | 2.2 | 6.8 | 1.4 | 0.6 | 2.4 | 47.7 | 1.3 | 3.3 | 4.3 | 1.0 | 1.5 | 0.0 | 3.3 |
| | Sardinia | 1.3 | 0.4 | 1.2 | 3.1 | 5.5 | 1.6 | 8.9 | 2.4 | 6.9 | 1.1 | 0.5 | 2.9 | 2.1 | 49.4 | 3.3 | 5.0 | 0.8 | 0.7 | 0.0 | 3.0 |
| | Sicily | 1.1 | 0.3 | 2.9 | 4.6 | 4.5 | 1.4 | 7.3 | 1.2 | 6.0 | 1.1 | 0.4 | 2.3 | 2.3 | 1.2 | 54.4 | 4.5 | 0.5 | 1.2 | 0.0 | 2.7 |
| | Tuscany | 1.3 | 0.3 | 1.0 | 4.7 | 6.8 | 2.0 | 8.4 | 2.2 | 7.6 | 1.1 | 0.5 | 2.8 | 2.4 | 1.4 | 2.9 | 48.6 | 1.4 | 1.2 | 0.0 | 3.6 |
| | Trentino Alto Adige | 0.7 | 0.1 | 0.8 | 2.3 | 5.1 | 2.5 | 5.4 | 1.5 | 7.7 | 1.0 | 0.3 | 3.2 | 1.5 | 0.6 | 2.0 | 5.8 | 52.1 | 1.0 | 0.1 | 6.3 |
| | Umbria | 1.9 | 0.8 | 1.4 | 4.1 | 5.8 | 2.1 | 9.1 | 1.3 | 6.9 | 1.7 | 0.2 | 2.4 | 2.8 | 0.8 | 3.1 | 5.1 | 1.1 | 46.7 | 0.1 | 2.7 |
| | Valle D'Aosta | 0.0 | 0.6 | 0.0 | 2.8 | 8.9 | 1.1 | 6.1 | 3.4 | 15.1 | 1.7 | 0.0 | 22.3 | 2.2 | 0.0 | 1.1 | 3.9 | 1.1 | 2.2 | 24.0 | 3.4 |
| | Veneto | 1.4 | 0.2 | 0.6 | 2.8 | 7.4 | 3.6 | 7.2 | 1.6 | 9.8 | 1.2 | 0.3 | 3.3 | 1.8 | 1.0 | 2.3 | 4.1 | 2.3 | 0.9 | 0.1 | 48.3 |

Figure 2 maps for each region the maximum values of inter-regional knowledge flows. When region A imports knowledge from region B more than from any other region, and exports knowledge to region B more than to any other region, the two regions are connected by a solid line, otherwise by a dotted line. As expected, most flows are bidirectional, and the largest regions, Lombardy and Latium, are pivotal. The geographic proximity effect is quite evident too, as Lombardy is pivotal to northern regions, and Latium to southern.

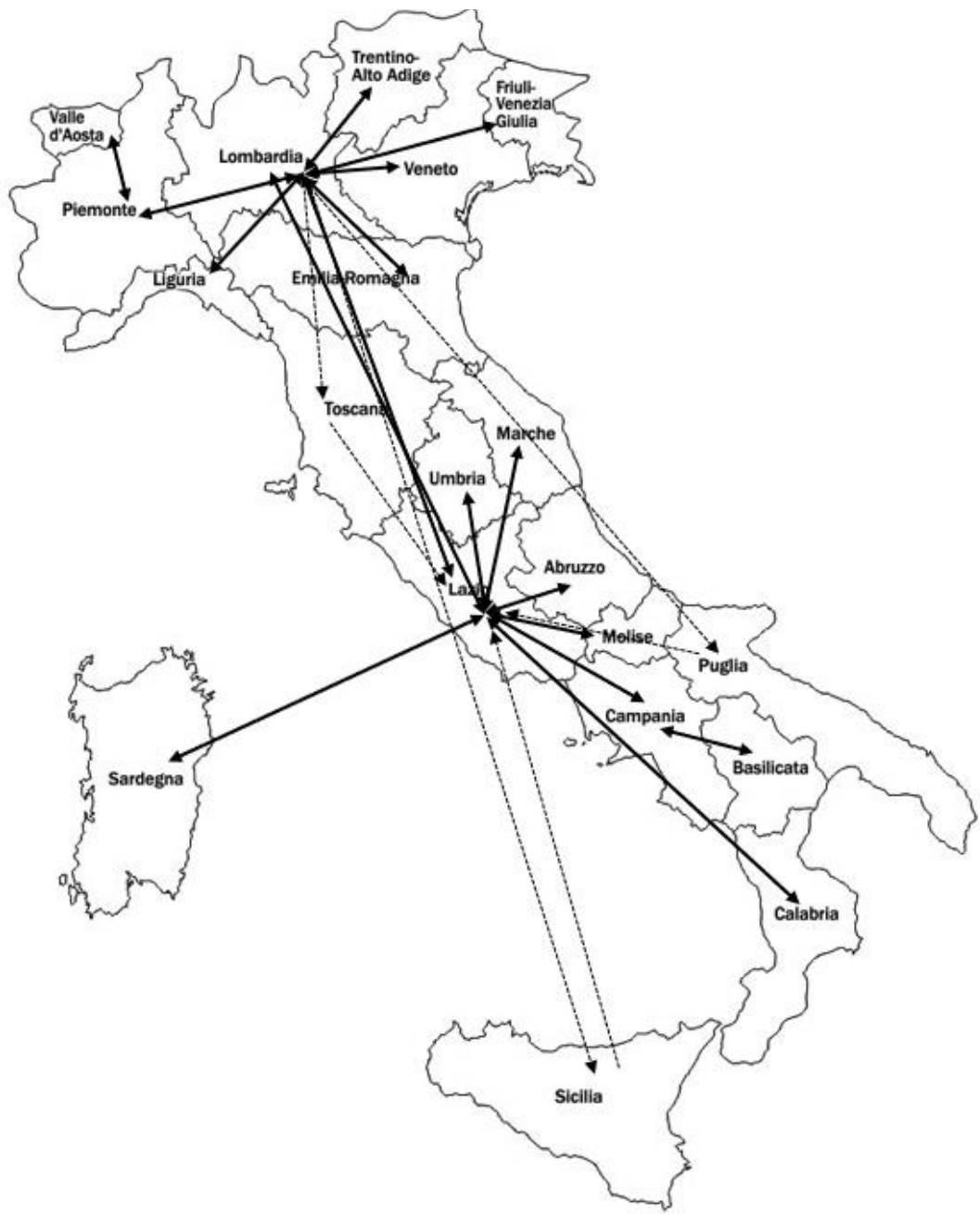

*Figure 2: Map of the maximum inter-regional knowledge flows (when region A imports knowledge from region B more than from any other region, and exports knowledge to region B more than to any other region, the two regions are connected by a solid line, otherwise by a dotted line)*

## 3.2 The RBKF at field level

To conduct a stratification of the RBKF at field level, we use the subject categories (SCs) of the WoS classification schema. In particular, we assign each cited publication to the SC of the hosting source (journal, conference, book, etc.). A "full counting" approach is adopted, meaning that a publication published in a multi-category journal is fully assigned to each SC associated to the journal. The cited publications of our dataset are distributed over 246 SCs, in turn grouped in 13 scientific macro-areas.[8]

As an example, Table 4 shows the value of the Tuscany RBKF in the SCs falling in the Biomedical research area. In this area, the "made in" Tuscany 2010-2012 publications generate altogether 17516 gains, 9421 of which extra-regional.

Vice versa, Tuscany earns 8492 gains from publications by the other Italian regions. The overall balance is therefore positive and equal to +929 units. By analysing data related to the single SCs, it can be noted that half of them show a positive balance, from a minimum of +33 in Medical laboratory technology to a max of +502 in Pharmacology & pharmacy; the remaining half show a nihil balance in Virology, and negative in six SCs (from -14 in Anatomy & morphology to -320 in Hematology).

*Table 4: Tuscany region balance of knowledge flows (RBKF) in the WoS subject categories falling in Biomedical research*

| Subject category | Extra-regional gains generated (a) | Earned gains (b) | RBKF (a-b) |
|---|---|---|---|
| Allergy | 108 | 123 | -15 |
| Anatomy & morphology | 19 | 33 | -14 |
| Chemistry, medicinal | 1391 | 977 | +414 |
| Hematology | 483 | 803 | -320 |
| Immunology | 1332 | 1079 | +253 |
| Infectious diseases | 240 | 426 | -186 |
| Medical laboratory technology | 153 | 120 | +33 |
| Medicine, research & experimental | 498 | 650 | -152 |
| Oncology | 1499 | 1465 | +34 |
| Pathology | 484 | 254 | +230 |
| Pharmacology & pharmacy | 1989 | 1487 | +502 |
| Radiology, nuclear medicine & medical imaging | 760 | 576 | +184 |
| Toxicology | 313 | 347 | -34 |
| Virology | 152 | 152 | 0 |
| Total | 9421 | 8492 | +929 |

When extending the analysis to all areas, SCs with a higher inclination to export (or import) of new knowledge may be identified. Table 5 reports the case of the first 10 SCs registering the lowest RBKF and the highest RBKF values for the Veneto region. SCs highly inclined to import, with a RBKF value ranging between -606 (Oncology) and -120 (Physics, multidisciplinary), are top of the list.

Actually, the prevailing presence of Bio-Med sciences SCs, with three SCs falling in Clinical Medicine and four in Biomedical Research, is quite evident.

---

[8] Mathematics; Physics; Chemistry; Earth and Space Sciences; Biology; Biomedical Research; Clinical Medicine; Psychology; Engineering; Economics; Law, political and social sciences; Art and Humanities; Multidisciplinary Sciences. The macro-areas and the assignment of SCs to them were at some point defined by the Institute of Scientific Information (ISI), although no longer showing in current Clarivate Analytics bibliographic products. There is no multi-assignment of SCs to macro-areas.

Less evident is the disciplinary concentration in the lower section of the table, with the only thing deserving attention being the presence of two SCs falling in Biology at the very bottom.

*Table 5: Subject categories with the highest and the lowest balance of knowledge flows (RBKF), for the Veneto region*

| Subject category | Area | Extra-regional gains generated (a) | Earned gains (b) | RBKF (a-b) |
|---|---|---|---|---|
| Oncology | Biomedical research | 921 | 1527 | -606 |
| Pharmacology & pharmacy | Biomedical research | 741 | 1056 | -315 |
| Gastroenterology & hepatology | Clinical medicine | 796 | 1059 | -263 |
| Physics, particles & fields | Physics | 751 | 980 | -229 |
| Cardiac & cardiovascular systems | Clinical medicine | 739 | 964 | -225 |
| Genetics & heredity | Clinical medicine | 266 | 470 | -204 |
| Radiology, nuclear medicine & medical imaging | Biomedical research | 527 | 715 | -188 |
| Geochemistry & geophysics | Earth and space science | 176 | 342 | -166 |
| Immunology | Biomedical research | 861 | 1005 | -144 |
| Physics, multidisciplinary | Physics | 191 | 311 | -120 |
| … | - | - | - | - |
| Electrochemistry | Chemistry | 232 | 126 | 106 |
| Energy & fuels | Physics | 382 | 243 | 139 |
| Rehabilitation | Clinical medicine | 267 | 123 | 144 |
| Agriculture, dairy & animal science | Biology | 285 | 133 | 152 |
| Hematology | Biomedical research | 961 | 804 | 157 |
| Surgery | Clinical medicine | 962 | 721 | 241 |
| Engineering, electrical & electronic | Engineering | 794 | 510 | 284 |
| Clinical neurology | Clinical medicine | 1361 | 967 | 394 |
| Biochemistry & molecular biology | Biology | 1753 | 1230 | 523 |
| Cell biology | Biology | 1425 | 773 | 652 |

The analysis of knowledge flows may also be carried out on pairs of regions in order to identify the SCs with the highest surplus or deficit in the bilateral relations between the two regional research systems considered. Table 6, for example, reports the analysis on the flows between Latium and Piedmont for the SCs falling in Earth and space science. The reported value of balances shows surplus and deficit by SC, from the Latium perspective. Overall, Latium exports more than it imports from Piedmont (308 vs 306) mainly by virtue of the flows generated by publications in Environmental sciences, in Geochemistry & geophysics and in Meteorology & atmospheric sciences. In the opposite direction the Latium RBKF is negative in only four SCs, and mainly in Geosciences, multidisciplinary (-38) and in Geology (-27).

Table 7 shows the extension to all areas of the above mentioned analysis, with reference to the bi-directional flows between Lombardy and Emilia Romagna, in order to identify the SCs showing the greatest spread between knowledge inflows and outflows from one region to the other. Table 7 takes the Lombardy perspective in determining the RBKF deficit or surplus. It is the other way around from the Emilia Romagna perspective. The upper part of the table shows not so important differences between SCs with the greatest decifit for Lombardy, while there is an evident imbalance of flows in the lower part of the table, with a substantial surplus of flows from Lombardy to Emilia Romagna especially in Physics, particles & fields (432) and in Astronomy & astrophysics (296).

*Table 6: The Latium – Piedmont regions balance of knowledge flows (RBKF) in the WoS subject categories of Earth and space science*

| Subject category | Gains from Latium to Piedmont (b) | Gains from Piedmont to Latium (a) | RBKF (a-b) |
| --- | --- | --- | --- |
| Geosciences, multidisciplinary | 59 | 97 | -38 |
| Geology | 7 | 34 | -27 |
| Paleontology | 9 | 22 | -13 |
| Green & sustainable science & technology | 2 | 10 | -8 |
| Geography, physical | 18 | 19 | -1 |
| Oceanography | 2 | 1 | +1 |
| Limnology | 3 | 1 | +2 |
| Water resources | 14 | 11 | +3 |
| Mineralogy | 8 | 1 | +7 |
| Environmental studies | 9 | 1 | +8 |
| Environmental sciences | 83 | 63 | +20 |
| Geochemistry & geophysics | 61 | 37 | +24 |
| Meteorology & atmospheric sciences | 33 | 9 | +24 |

*Table 7: The Lombardy – Emilia Romagna regions balance of knowledge flows (RBKF) in the bottom and top 10 WoS subject categories per Lombardy RBKF deficit and surplus*

| Subject category | Area* | Gains from Lombardy to Emilia Rom. | Gains from Emilia Rom. to Lombardy | RBKF |
| --- | --- | --- | --- | --- |
| Dentistry, oral surgery & medicine | 7 | 62 | 146 | -84 |
| Microbiology | 5 | 95 | 171 | -76 |
| Engineering, biomedical | 9 | 116 | 187 | -71 |
| Obstetrics & gynecology | 7 | 104 | 167 | -63 |
| Chemistry, multidisciplinary | 3 | 195 | 256 | -61 |
| Orthopedics | 7 | 77 | 134 | -57 |
| Reproductive biology | 5 | 73 | 126 | -53 |
| Chemistry, medicinal | 6 | 67 | 113 | -46 |
| Hematology | 6 | 326 | 370 | -44 |
| Energy & fuels | 2 | 35 | 72 | -37 |
| … | - | - | - | - |
| Environmental sciences | 4 | 308 | 233 | +75 |
| Instruments & instrumentation | 9 | 121 | 45 | +76 |
| Urology & nephrology | 7 | 126 | 46 | +80 |
| Cell biology | 5 | 278 | 187 | +91 |
| Neurosciences | 7 | 556 | 463 | +93 |
| Clinical neurology | 7 | 356 | 247 | +109 |
| Medicine, research & experimental | 6 | 172 | 61 | +111 |
| Oncology | 6 | 558 | 403 | +155 |
| Astronomy & astrophysics | 2 | 573 | 277 | +296 |
| Physics, particles & fields | 2 | 496 | 64 | +432 |

\* 1, Mathematics; 2, Physics; 3, Chemistry; 4, Earth and Space Sciences; 5, Biology; 6, Biomedical Research; 7, Clinical Medicine; 8, Psychology; 9, Engineering.

### 3.3 The knowledge flows specialization indexes

In this subsection, we measure the specialization indexes for outflows and inflows of knowledge by a given region. In simple terms, they measure the extent to which a region's knowledge flows differ from those of the rest of the country or a comparison group of regions. The relevant indicators are the "knowledge outflows specialization index" (KOSI) and the "knowledge inflows specialization index" (KISI). They measure respectively a region's capacity to "export" knowledge to other regions, or to "import" knowledge from other regions, as compared to the rest of the country, across all research fields. In operational terms, KOSI is calculated here applying the "revealed comparative advantage" (RCA) methodology and, in particular, the *Balassa index* (Balassa, 1979).

The KOSI and KISI of country k in the SCj (respectively $KOSI_{kj}$ and $KISI_{kj}$) are defined as:

$$KOSI_{kj} = 100 * \tanh ln \left\{ \frac{(G_{kj}/\sum_{i \neq j} G_{ki})}{\sum_{z \neq k} G_{zj} / \sum_{z \neq k} \sum_{i \neq j} G_{zi}} \right\}$$

and

$$KISI_{kj} = 100 * \tanh ln \left\{ \frac{(G_{kj}/\sum_{i \neq j} G_{ki})}{\sum_{z \neq k} G_{zj} / \sum_{z \neq k} \sum_{i \neq j} G_{zi}} \right\}$$

with $G_{kj}$ indicating the gains generated (KOSI) or earned (KISI) by region *k* in the SC*j*.

Use of the logarithmic function centers the data around zero and the hyperbolic tangent multiplied by 100 limits the $KOSI_{kj}$ and $KISI_{kj}$ values to a range of +100 to -100. For any SC, the closer the value of the index to +100 the more the region is specialized in that SC in generating (appropriating) knowledge flows to (from) other regions. Vice versa, the closer the index approaches -100, the less the region is specialized in the SC. Values around 0 are labeled as "expected".

Tables 8-10 show the results of the application of the specialization index for the outflows of knowledge, listing the ten SCs with the highest values of $KOSI_{kj}$, for each region. Similarly, Tables 11-13 list the ten SCs with the highest value of $KISI_{kj}$, for the inflows of knowledge in each region.

The analysis clearly shows the potential of this tool which illustrates the fields where a region is relatively more specialized in exporting or importing new knowledge to and from other regions.

From a different perspective, the two indicators can be used to identify, in a given field, which regions are more specialized in terms of knowledge outflows and inflows. To this purpose, Figure 3 shows the regions with the highest and lowest value of KISI for the 11 SC of Agricultural sciences, while Figure 4 shows the regions with the highest and lowest value of KISI, for the same fields.

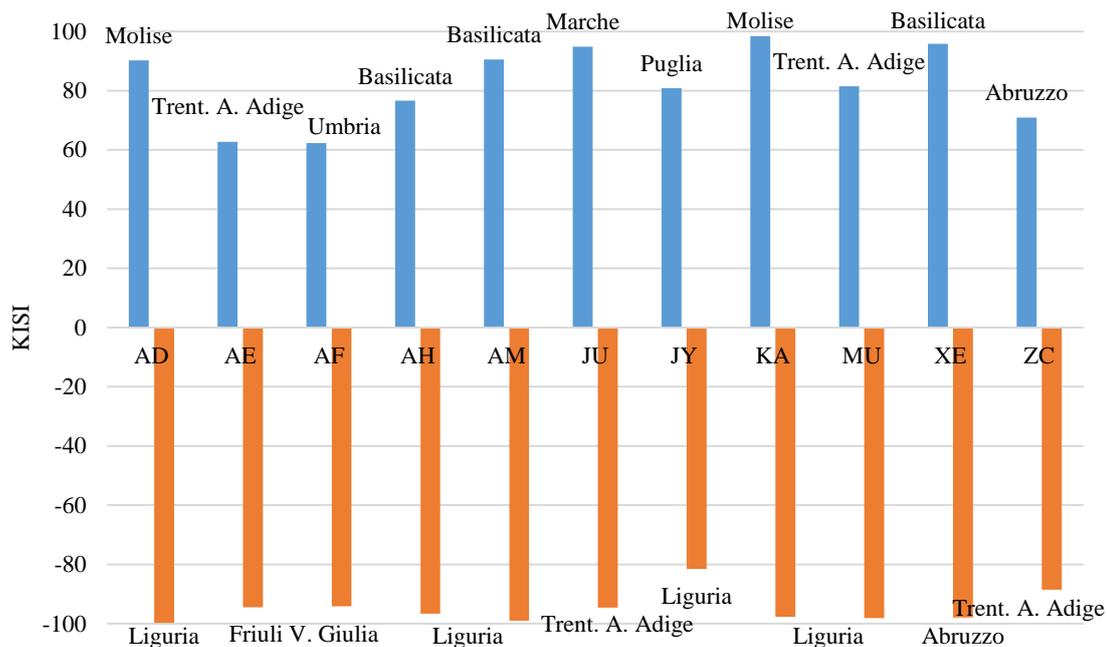

*Figure 3: Regions with the highest and lowest knowledge inflows specialization indexes (KISI) in each SC of the Agricultural sciences area (AD-Agriculture, dairy & animal science; AE-Agricultural engineering; AF-Agricultural economics & policy; AH-Agriculture, multidisciplinary; AM-Agronomy; JU-Fisheries; JY-Food science & technology; KA-Forestry; MU-Horticulture; XE-Soil science; ZC-Veterinary sciences)*

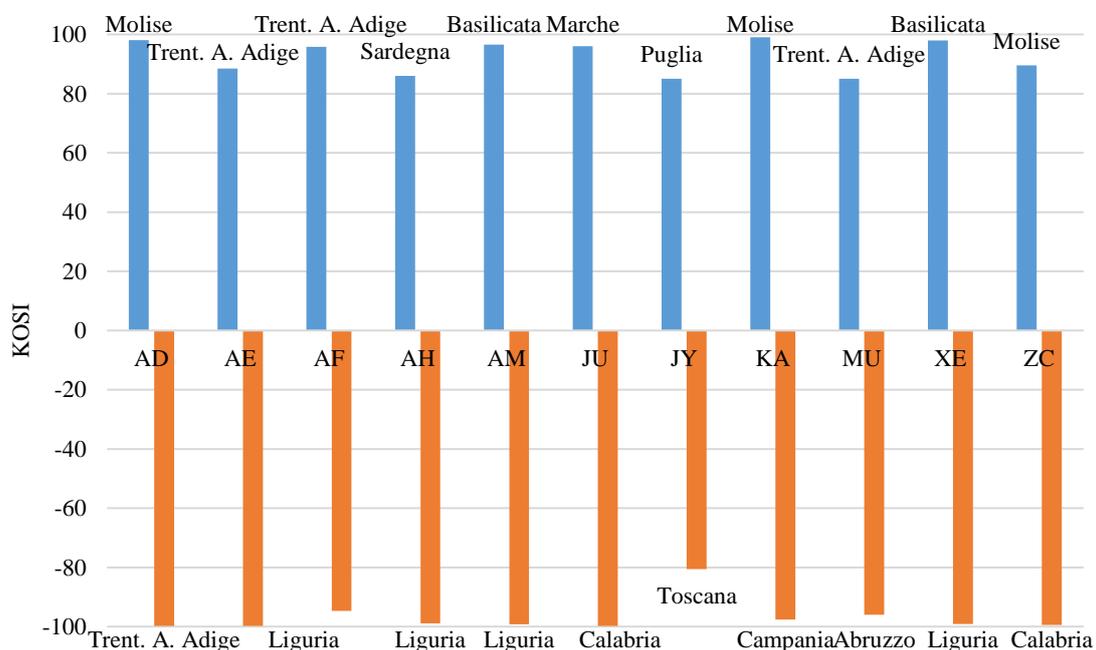

*Figure 4: Regions with the highest and lowest knowledge ouflows specialization indexes (KOSI) in each SC of the Agricultural sciences area (AD-Agriculture, dairy & animal science; AE-Agricultural engineering; AF-Agricultural economics & policy; AH-Agriculture, multidisciplinary; AM-Agronomy; JU-Fisheries; JY-Food science & technology; KA-Forestry; MU-Horticulture; XE-Soil science; ZC-Veterinary sciences)*

*Table 8: Subject categories with the highest knowledge outflows specialization indexes, KOSI, (values in brackets) by regions in North of Italy*

| Emilia Romagna | Friuli Venezia Giulia | Liguria | Lombardia | Piemonte | Trentino Alto Adige | Valle D'Aosta | Veneto |
|---|---|---|---|---|---|---|---|
| Womens studies (94.6) | Film, radio, television (99.8) | Robotics (99.1) | Cultural studies (97.9) | Materials science, textiles (99.7) | Ornithology (97.4) | Art (99.9) | Psychology, educational (97.4) |
| Orthopedics (91.7) | Oceanography (96.7) | Engineering, marine (98.6) | Religion (97.3) | Literature (99.2) | Telecommunications (96.7) | Business, finance (99.8) | Agriculture, dairy & animal science (89.6) |
| Materials science, ceramics (90.2) | Communication (95.9) | Engineering, ocean and marine (96.1) | Film, radio, television (91.7) | Materials science, textiles, paper & wood (97.7) | Materials science, textiles, paper & wood (96.7) | Transportation (99.7) | Medicine, legal (88.5) |
| History of social sciences (89.2) | Logic (95.2) | Allergy (93.0) | Literature (91.7) | Cultural studies (92.8) | Imaging science & photographic technology (96.7) | Mathematical & computational biology (99.6) | Medical laboratory technology (87.4) |
| Medical laboratory technology (85.8) | Nursing (92.8) | Microscopy (90.2) | Classics (87.3) | Social work (89.1) | Language & linguistics (96.6) | Statistics & probability (99.6) | Social work (83.6) |
| History (83.9) | Engineering, petroleum (92.1) | Classics (89.3) | Public administration (87.1) | Engineering, aerospace (88.8) | Agricultural economics & policy (95.8) | Radiology, nuclear medicine & medical imaging (99.5) | Limnology (82.5) |
| Medieval & renaissance studies (81.8) | Social issues, multidisciplinary (87.9) | Rheumatology (87.4) | Philosophy (86.3) | History & philosophy of science (88.7) | Forestry (95.8) | Zoology (99.4) | Education, special (82.3) |
| Psychology, clinical (80.2) | Physics, multidisciplinary (86.5) | Computer science, artificial intelligence (87.2) | Business (84.7) | Multidisciplinary sciences (79.4) | Remote sensing (95.8) | Archaeology (99.1) | Psychology, biological (82.1) |
| Classics (76.1) | Astronomy & astrophysics (84.2) | Marine & freshwater biology (85.5) | Communication (81.3) | Mining & mineral processing (78.2) | Materials science, ceramics (94.6) | Computer science, interdisciplinary applications (97.6) | Psychology, multidisciplinary (78.6) |
| Emergency medicine (74.4) | Audiology & speech-language pathology (83.9) | Oceanography (80.9) | Cell & tissue engineering (80.1) | Computer science, hardware & architecture (77.3) | Sociology (94.0) | Oncology (95.9) | Psychology, social (78.5) |

*Table 9: Subject categories with the highest knowledge outflows specialization indexes, KOSI, (values in brackets) by regions in Center of Italy*

| Abruzzo | Lazio | Marche | Toscana | Umbria |
|---|---|---|---|---|
| Psychology, mathematical (98.2) | Primary health care (99.9) | Humanities, multidisciplinary (99.1) | Literary theory & criticism (99.4) | Medical ethics (99.4) |
| Hospitality, leisure, sport & tourism (97.3) | Psychology, psychoanalysis (93.8) | International relations (98.1) | Andrology (99.2) | Mining & mineral processing (95.0) |
| Industrial relations & labor (96.0) | Information science & library science (93.0) | Marine & freshwater biology (96.3) | Literature, german, dutch, scandinavian (91.3) | Humanities, multidisciplinary (94.8) |
| Integrative & complementary medicine (93.8) | Medieval & renaissance studies (84.5) | Fisheries (96.1) | Ethnic studies (89.6) | Physics, atomic, molecular & chemical (92.1) |
| Physiology (93.8) | Social sciences, biomedical (82.0) | Business, finance (95.9) | Psychology, applied (88.3) | Medicine, general & internal (92.1) |
| Parasitology (93.4) | Engineering, marine (77.7) | Art (95.4) | Robotics (85.9) | Limnology (85.0) |
| History & philosophy of science (92.9) | Sport sciences (77.1) | Microscopy (95.4) | Rheumatology (85.4) | Chemistry, physical (83.8) |
| Dentistry, oral surgery & medicine (91.9) | Medical ethics (75.2) | Construction & building technology (94.4) | Mining & mineral processing (81.4) | Water resources (83.6) |
| Veterinary sciences (89.5) | Substance abuse (73.7) | Tropical medicine (93.5) | Sociology (78.4) | Green & sustainable science & technology (82.2) |
| Psychology, biological (88.8) | Geochemistry & geophysics (72.7) | Mycology (92.8) | Art (77.5) | Mycology (81.3) |

*Table 10: Subject categories with the highest knowledge outflows specialization indexes, KOSI, (values in brackets) by regions in South of Italy*

| Basilicata | Calabria | Campania | Molise | Puglia | Sardegna | Sicilia |
|---|---|---|---|---|---|---|
| Area studies (100.0) | Transportation (96.1) | Dance, theater, music, film and folklore (98.5) | Forestry (99.1) | Tropical medicine (95.8) | Literature, german, dutch, scandinavian (99.8) | Ethnic studies (97.7) |
| Materials science, textiles, paper & wood (99.8) | Logic (93.4) | Criminology & penology (91.7) | Agriculture, dairy & animal science (98.1) | Crystallography (94.3) | Substance abuse (96.3) | Psychology, psychoanalysis (96.2) |
| Classics (99.7) | Engineering, marine (93.4) | Engineering, petroleum (87.7) | International relations (96.7) | Parasitology (93.9) | Chemistry, inorganic & nuclear (96.1) | Family studies (96.1) |
| Engineering, geological (99.0) | Engineering, petroleum (93.4) | Materials science, composites (86.8) | Medicine, general & internal (94.8) | Agronomy (89.7) | Hospitality, leisure, sport & tourism (95.9) | Physics, nuclear (90.8) |
| Soil science (98.0) | Engineering, ocean and marine (89.8) | Engineering, geological (84.1) | Nutrition & dietetics (92.6) | Education, special (88.7) | Integrative & complementary medicine (93.6) | Agricultural engineering (78.9) |
| Agronomy (96.6) | Art (89.3) | Transportation (82.7) | Psychology, biological (90.1) | Ethics (85.3) | Demography (92.4) | Materials science, coatings & films (78.5) |
| Integrative & complementary medicine (96.2) | Archaeology (89.1) | Engineering, chemical (81.7) | Agronomy (90.1) | Food science & technology (85.0) | Nursing (92.4) | Green & sustainable science & technology (70.5) |
| Urban studies (95.3) | Transportation science & technology (86.8) | Thermodynamics (78.0) | Ecology (89.9) | Horticulture (84.9) | Education, scientific disciplines (91.4) | Nuclear science & technology (68.8) |
| Remote sensing (94.7) | Water resources (80.8) | Ergonomics (77.0) | Veterinary sciences (89.5) | Microscopy (84.2) | Anthropology (90.2) | Marine & freshwater biology (67.6) |
| Geosciences, multidisciplinary (93.4) | Computer science, software engineering (79.2) | Polymer science (75.0) | Evolutionary biology (89.0) | Agriculture, dairy & animal science (81.0) | Obstetrics & gynecology (88.3) | Geology (67.1) |

*Table 11: Subject categories with the highest knowledge inflows specialization indexes, KISI, (values in brackets) by regions in North of Italy*

| Emilia Romagna | Friuli Venezia Giulia | Liguria | Lombardia | Piemonte | Trentino Alto Adige | Valle D'Aosta | Veneto |
|---|---|---|---|---|---|---|---|
| Orthopedics (83.8) | Film, radio, television (98.4) | Robotics (98.4) | Classics (95.9) | Materials science, textiles (99.2) | Area studies (96.8) | Religion (99.9) | Psychology, educational (93.9) |
| Materials science, ceramics (82.8) | Nursing (92.9) | Engineering, marine (98.2) | Literature (93.6) | Literature (99.2) | History & philosophy of science (96.3) | Social work (99.9) | Social work (91.3) |
| Medical laboratory technology (81.4) | Medieval & renaissance studies (89.8) | Classics (97.3) | Film, radio, television (86.2) | Materials science, textiles, paper & wood (97.4) | Ornithology (95.8) | Hospitality, leisure, sport & tourism (99.6) | Medical laboratory technology (85.7) |
| Medieval & renaissance studies (81.4) | Engineering, petroleum (87.6) | Engineering, ocean and marine (93.0) | Communication (85.5) | Cultural studies (97.0) | Hospitality, leisure, sport & tourism (95.1) | Transportation (98.8) | Family studies (85.0) |
| Womens studies (79.2) | Oceanography (86.1) | Allergy (88.7) | Ornithology (79.1) | Literature, german, dutch, scandinavian (97.0) | Imaging science & photographic technology (94.9) | Limnology (98.4) | Agriculture, dairy & animal science (82.7) |
| Emergency medicine (77.7) | Astronomy & astrophysics (84.8) | Computer science, artificial intelligence (86.6) | Public administration (79.1) | Engineering, aerospace (87.3) | Remote sensing (94.2) | Mining & mineral processing (98.3) | Psychology, biological (80.6) |
| Andrology (73.1) | Logic (84.7) | Microscopy (84.1) | Social issues, multidisciplinary (77.9) | History & philosophy of science (82.5) | Social issues, multidisciplinary (93.9) | History (97.5) | Medicine, legal (80.6) |
| Materials science, characterization & testing (58.2) | Physics, particles & fields (83.3) | Computer science, cybernetics (79.9) | Medical ethics (73.2) | Humanities, multidisciplinary (82.5) | Materials science, textiles, paper & wood (93.2) | Art (96.9) | Psychology, mathematical (79.5) |
| Cultural studies (55.7) | Physics, multidisciplinary (80.4) | Rheumatology (77.0) | Business (72.3) | Religion (80.9) | Forestry (93.1) | Radiology, nuclear medicine & medical imaging (96.3) | Limnology (74.3) |
| Anatomy & morphology (52.5) | Communication (79.8) | Oceanography (75.2) | Ethics (69.1) | Social work (79.2) | Limnology (93.0) | Forestry (96.2) | Psychology, developmental (73.4) |

*Table 12: Subject categories with the highest knowledge inflows specialization indexes, KISI, (values in brackets) by regions in Center of Italy*

| Abruzzo | Lazio | Marche | Toscana | Umbria |
|---|---|---|---|---|
| Industrial relations & labor (91.3) | Primary health care (95.3) | Humanities, multidisciplinary (99.0) | Literature, german, dutch, scandinavian (99.5) | Physics, atomic, molecular & chemical (82.9) |
| Hospitality, leisure, sport & tourism (89.9) | Information science & library science (90.8) | Fisheries (94.9) | Cultural studies (92.3) | Industrial relations & labor (82.1) |
| Physiology (87.6) | Psychology, psychoanalysis (82.8) | Marine & freshwater biology (94.0) | Psychology, applied (88.0) | Mycology (75.4) |
| Pathology (83.5) | Engineering, marine (82.6) | Engineering, petroleum (91.9) | Andrology (84.3) | Pathology (75.2) |
| Dentistry, oral surgery & medicine (81.0) | Medical ethics (77.4) | Womens studies (88.4) | History (78.5) | Green & sustainable science & technology (75.0) |
| Psychology, mathematical (80.9) | Education, special (66.8) | International relations (88.4) | Forestry (78.4) | Immunology (73.7) |
| Parasitology (79.8) | Social sciences, biomedical (64.7) | Gerontology (87.0) | International relations (75.8) | Fisheries (72.5) |
| Immunology (78.1) | Sport sciences (62.7) | Education, special (86.4) | History of social sciences (74.6) | Physiology (71.7) |
| Philosophy (75.8) | Geography (62.2) | Microscopy (85.2) | Physics, atomic, molecular & chemical (72.7) | Construction & building technology (70.7) |
| Psychiatry (75.6) | Orthopedics (60.1) | Construction & building technology (83.8) | Sociology (72.0) | Energy & fuels (68.4) |

*Table 13: Subject categories with the highest knowledge inflows specialization indexes, KISI, (values in brackets) by regions in South of Italy*

| Basilicata | Calabria | Campania | Molise | Puglia | Sardegna | Sicilia |
|---|---|---|---|---|---|---|
| Area studies (99.9) | Transportation (94.2) | Dance, theater, music, film and folklore (98.5) | Forestry (98.5) | Psychology, psychoanalysis (89.0) | Substance abuse (96.7) | Ethnic studies (92.8) |
| Classics (99.9) | Engineering, marine (87.6) | Criminology & penology (91.8) | Anesthesiology (96.8) | Parasitology (87.8) | Criminology & penology (95.6) | Psychology, psychoanalysis (91.8) |
| Materials science, textiles, paper & wood (98.0) | Engineering, petroleum (85.4) | Film, radio, television (89.3) | Cell & tissue engineering (93.5) | Crystallography (86.0) | Chemistry, inorganic & nuclear (93.1) | Family studies (89.4) |
| Soil science (95.9) | Area studies (85.0) | Materials science, composites (81.9) | Agriculture, dairy & animal science (90.2) | Tropical medicine (85.9) | Hospitality, leisure, sport & tourism (92.6) | Physics, nuclear (80.7) |
| Engineering, geological (95.6) | Engineering, ocean and marine (83.3) | Primary health care (81.6) | Critical care medicine (89.8) | Ethnic studies (81.5) | Anthropology (86.0) | Materials science, coatings & films (73.3) |
| Geography (92.8) | Transportation science & technology (81.0) | Transportation (78.0) | Urban studies (88.9) | Dance, theater, music, film and folklore (81.5) | Agriculture, dairy & animal science (84.2) | Fisheries (59.0) |
| Remote sensing (91.5) | Logic (80.0) | Thermodynamics (77.1) | Ophthalmology (85.4) | Food science & technology (80.9) | Biodiversity conservation (83.1) | Marine & freshwater biology (58.3) |
| Mining & mineral processing (91.3) | Mathematics, applied (78.8) | Engineering, chemical (76.3) | Radiology, nuclear medicine & medical imaging (84.3) | Agronomy (79.4) | Logic (82.5) | Nuclear science & technology (56.9) |
| Agronomy (90.5) | Computer science, software engineering (73.1) | Transportation science & technology (73.7) | Plant sciences (83.9) | Horticulture (78.8) | Demography (81.9) | Thermodynamics (56.3) |
| Geosciences, multidisciplinary (89.2) | Engineering, chemical (71.3) | Area studies (71.6) | Ecology (81.9) | Microscopy (77.0) | Obstetrics & gynecology (76.8) | Horticulture (54.2) |

# 4. Conclusions

This work has applied a new approach to measure knowledge flows, based on linkages between cited publications "made in" a given region and citing publications "made in" other regions of the same country. In doing so, we have been able to construct a regional balance of knowledge flows, at both overall and field level. Furthermore, we calculated for each region its capacity to "export" (import) knowledge to (from) other regions, as compared to all other regions of the country, across all research fields.

Compared to previous literature, the recourse to publication citations rather than patent citations offers a different perspective, and an order of magnitude of observations much higher than that of patents.

While results cannot be generalized to other countries, some emerging evidence could be of general interest. There occurs a positive and strong correlation between the size of scientific output of a region and the value of its RBKF. Larger regions are better able to export new knowledge. At the same time we registered a positive correlation between the share of intraregional gains and the size of overall scientific output of a region. This can be due to the fact that in large regions it is likely to find large research laboratories/groups conducting research on topics of common interest.

The study can be easily replicated in other national contexts, and provide useful information to the policy maker: at aggregate level it allows to measure the share of intra- vs extra-regional knowledge flows generated by a region's research system, compared to other regions. At field level, the RBKF allows to pinpoint the subject fields with a higher propensity to export (or import) new knowledge to (from) other regions. Moreover, its application on pairs of regions allows to identify the field with the highest surplus or deficit in the bilateral relations between the two regional research systems considered. The possible longitudinal analysis of the RBKF could support the assessment of the effectiveness of research policies undertaken over time.

The methodology developed provides useful results for informing both national and regional research policies, for example the analysis of comparative advantage of regions could be particularly pertinent concerning bilateral collaborations. Extending the observation period, would allow cross-time analysis to monitor how such comparative advantages vary along time.

Furthermore, it is generally the case that universities are the key loci of research in a country. Their production capacity is planned though mainly to satisfy the demand for higher education, rather than that for research. In such countries as Italy, characterized by low mobility of students, for cultural and economic reasons, less inhabited territories are disfavored in terms of research. Abramo, D'Angelo, and Di Costa (2020a) have confirmed that in Italy too geographic proximity favors knowledge flows, and showed how this effect varies across research fields (Abramo, D'Angelo, & Di Costa, 2020b). The question then is whether, due to the geographic proximity effect on domestic knowledge flows, regions that are strong in certain subject fields could have a support function towards other regions, less well developed in those fields.

A strictly connected stream of research would be investigating whether the geographic proximity effect tends to fade away over time. The authors have already started looking into that.